\documentclass[preprint,12pt,natbib]{elsarticle}

\usepackage{amssymb}
\usepackage{lipsum}
\usepackage{siunitx}
\newcommand{\activeliquid}{$C_2H_2F_4$}

\usepackage{hyperref}
\hypersetup{
	colorlinks=true,
	linkcolor=blue,
	filecolor=blue,      
	urlcolor=blue,
}

\journal{Astroparticle Physics}
\begin{document}
	
	\begin{frontmatter}
		
		
		
		\title{First result from tetrafluoroethane ($C_2H_2F_4$) superheated emulsion detector for dark matter search at JUSL}
		
		% All author list is written in author_list.tex file.
		%\input{author_list}
		
		\author[aff1,aff2]{V. Kumar}
		\affiliation[aff1]{{Saha Institute of Nuclear Physics},
			addressline={1/AF, Salt Lake}, 
			city={Kolkata},
			postcode={700064}, 
			state={West Bengal},
			country={India}}
		\affiliation[aff2]{{Homi Bhabha National Institute, Training School Complex},
			addressline={Anushakti Nagar}, 
			city={Mumbai},
			postcode={400094}, 
			state={Maharashtra},
			country={India}}
		
		\author[aff3]{S. Ali}
		\affiliation[aff3]{organization={Physics Department, Jadavpur University},
			addressline={Jadavpur},
			city={Kolkata},
			postcode={700032},
			state={West Bengal},
			country={India}}
		\author[aff1,aff2]{M. Das\corref{cor1} %\fnref{fn1}
		}
		\ead{mala.das@saha.ac.in}
		\cortext[cor1]{Corresponding author}
		%\fntext[fn1]{}
		
		\author[aff1]{N. Biswas}
		
		\author[aff1,aff2]{S. Das}
		
		\author[aff4]{S. Sahoo}
		\affiliation[aff4]{organization={Variable Energy Cyclotron Centre},
			addressline={1/AF, Salt Lake},
			city={Kolkata},
			postcode={700064},
			state={West Bengal},
			country={India}}
		
		\author[aff4]{N. Chaddha}
		
		\author[aff1]{J. Basu}
		
		\author[aff5]{V. N. Jha}
		\affiliation[aff5]{organization={Health Physics Unit, BARC},
			city={Jaduguda},
			postcode={832102},
			state={Jharkhand},
			country={India}}
		
		\begin{abstract}
			The superheated emulsion detector consisting of the droplets of tetra-fluoro-ethane (\activeliquid) has been fabricated at the laboratory and installed at the 555m deep underground laboratory, JUSL during July to Dec 2022. The 500ml detector ran for an effective period of 48.6 days at a threshold of 5.87 keV with an exposure of 2.47 kg-days. The acoustic signals produced due to the bubble nucleation were collected by the acoustic sensor and FPGA-based data acquisition system. The data shows a minimum sensitivity of \texorpdfstring{[$7.834\pm(0.370)_{statistical}(^{+2.005}_{-1.241})_{systematic}]\times 10^{-38}$ $cm^2$} for SI-nucleon for carbon at WIMP mass of 22.81 $GeV/c^2$ and \texorpdfstring{[$3.782\pm(0.179)_{statistical}(^{+0.655}_{-0.432})_{systematic}]\times 10^{-36}$ $cm^2$} for SD (p) for fluorine at 30.67 $GeV/c^2$.The threshold of WIMP mass is 5.16 $GeV/c^2$ for F and 4.44 $GeV/c^2$ for C at the operating threshold of 5.87 keV. The first result of the dark matter direct search experiment named InDEx with tetra-fluoro-ethane active liquid from JUSL underground laboratory is reported in this article. 
			
		\end{abstract}
		
		
		
	\end{frontmatter}
	
	%\tableofcontents
	
	
	
	\section{Introduction}
	\label{introduction}
	Astronomical observations and measurements of cosmic microwave background confirm the existence of dark matter (DM) but the particle nature and its interaction remain unknown \cite{RevModPhys.90.045002, Clowe_2006}. A most favorable candidate of DM named Weakly Interacting Massive Particles (WIMPs) are predicted in many theories beyond the standard model that may be responsible for the observed relic density \cite{PhysRevD.31.3059,JUNGMAN1996195}. The direct DM search experiments mainly look for the nuclear recoil signals arising from the WIMP-nucleus elastic scattering by using different detector technologies. The world-leading direct detection experiments are mostly sensitive in 20-30 $GeV/c^2$ WIMP mass region and the null results from those experiments have pushed the interest to explore the low WIMP mass region, especially below 20 $GeV/c^2$ \cite{Schumann_2019}. The detector to be sensitive in the low WIMP mass requires low threshold energy and a target containing low mass nuclei. Due to its weak interaction with the normal matter, the expected event rate from WIMP is extremely low. The experiments are looking for the signal coming from the spin-independent WIMP-nucleon interactions and WIMP-neutron/proton spin-dependent interactions. In the low mass WIMP search, DarkSide-50 experiment puts stringent exclusion limit below 5 $GeV/c^2$ WIMP mass with a cross-section of $10^{-40} cm^2$ and CRESST-III experiment showed the sensitivity below 1.6 $GeV/c^2$ WIMP mass with minimum sensitivity of the order of $10^{-36} cm^2$ at 0.5 $GeV/c^2$ \cite{PhysRevLett.121.081307,PhysRevD.100.102002,PhysRevD.107.122003}. Other experiments like NEWS-G, Super-CDMS, CDEX have started to explore the low WIMP mass region below 10 $GeV/c^2$ \cite{ARNAUD201854,PhysRevD.99.062001,PhysRevLett.120.241301}. In the spin-dependent sector, the PICASSO experiment provides limits in the WIMP mass region of 2 $GeV/c^2$ and 4 $GeV/c^2$ with a cross-section of $10^{-37} cm^2$ \cite{BEHNKE201785}. Few experiments have also started to venture the WIMP-electron scattering and one of such experiments, SENSEI puts constraints in the 0.5-5 $MeV/c^2$ mass range for the WIMP-electron scattering that produces electron recoil induced signal \cite{PhysRevLett.122.161801}. However, in the present work, we are focusing on the WIMP-nucleus scattering. 
	
	\par
	
	Detector operated for WIMPs search requires a reduced background environment for the operation. The neutrons, gamma rays, and alpha particles are known as major backgrounds. The experiments are operated deep underground to reduce the background arising from cosmic rays \cite{Aubin_2008}. Superheated liquid as the target material and as detector provides an excellent rejection to the backgrounds by adjusting the operating temperature and pressure of the liquid. The superheated liquid in the form of micron-sized droplets known as superheated emulsion detector (SED) has been used for a long time in various research fields where each droplet acts as a tiny bubble chamber \cite{DAS2005570,https://doi.org/10.1118/1.598844,ING199473}. SED is continuously sensitive to the energetic particles and does not need to be pressurized after each cycle of the bubble nucleation.
	
	\par
	
	The bubble nucleation in the SED occurs if the energy deposited by the incoming particle is equal to or greater than the critical energy of the liquid at a given temperature and pressure. SED generates acoustic pulses during the bubble nucleation that can be detected by acoustic sensors like piezoelectric transducers or microphones. The parameters can be constructed from the power and frequency of the acoustic pulses which helps to discriminate the events originating from low and high linear energy transfer (LET) particles \cite{MONDAL2013182, SETH201692}. The sensitive region of the SED can be changed by choosing a suitable threshold energy of the detector as at a particular superheat region it would be sensitive to high LET particles like neutrons and insensitive to low LET radiations like gamma-rays, beta particles, etc \cite{DERRICO2001229,SAHOO201944}. All these properties made SED to be used extensively in the WIMP search experiment \cite{PhysRevD.100.022001,PhysRevD.89.072013}. 
	
	\par
	
	In the present work, the measurement has been carried out at 555m deep underground at the Jaduguda Underground Science Laboratory (JUSL), Jaduguda Mine, Jharkhand, India. The experiment, named \textbf{InDEx} (\textbf{\underline{In}}dian \textbf{\underline{D}}ark matter search \textbf{\underline{Ex}}periment) at JUSL was carried out with {\activeliquid} (b.p. $-26.3^{\circ}C$) SED and the earlier theoretical study shows that {\activeliquid} SED would be sensitive to low WIMP mass down to sub-GeV while operated at low threshold \cite{PhysRevD.101.103005}. The noise level captured by the acoustic sensors has already been studied \cite{SAHOO2021165450} and the background study for the cosmic muons, neutrons, and gamma-rays have been performed using plastic scintillator detector, pressurized $^4He$ detector and CsI respectively and simulation has also been carried out \cite{SHARAN2021165083,GHOSH2022102700}. The present experiment has been done with {\activeliquid} SED with 2.47 kg-days of exposures at a 5.87 keV threshold. It is already shown that the hydrogen nuclei of the active liquid are not sensitive at this threshold for the bubble nucleation to occur \cite{PhysRevD.101.103005}. The minimum WIMP mass sensitive to the detector at this threshold and the WIMP-nucleon spin-independent and spin-dependent cross-section has been estimated for the carbon and fluorine nuclei.
	
	\section{Experiment}
	\label{experiment}
\begin{figure}[htbp]
		\centering
		\includegraphics[width=0.9\textwidth]{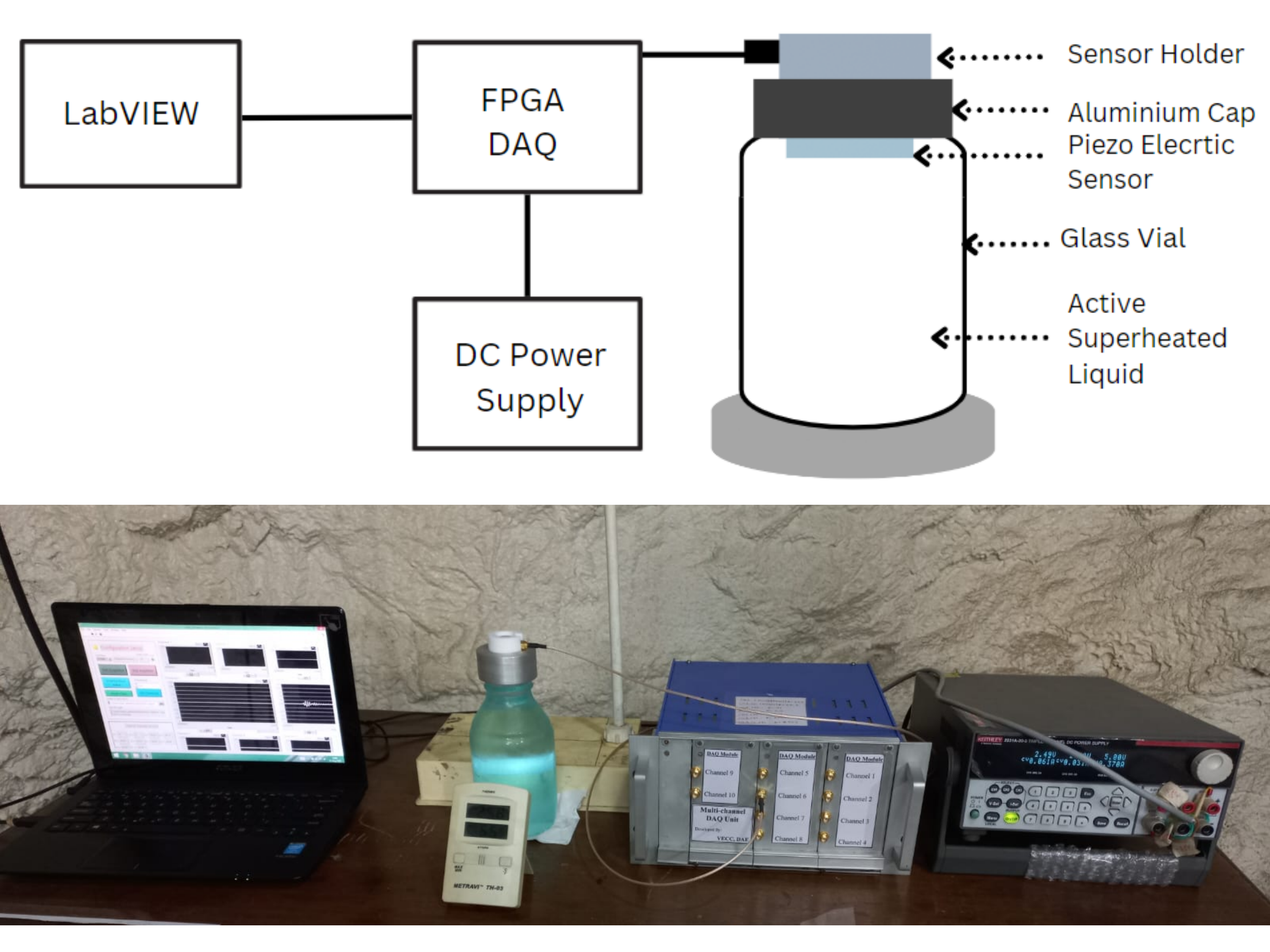}
		\caption{The experimental set-up} 
		\label{fig1}
	\end{figure}

	The superheated emulsion detector of 500ml was fabricated at the laboratory in a pressure reactor (maker: Amar Equipments Pvt. Ltd) by making the droplets of superheated liquid in a gel matrix. The detector was installed at JUSL at the room temperature of average $25.4^{\circ}C$ along with the piezoelectric acoustic sensor coupled at the top of the detector touching the aquasonic gel. The output from the sensor was connected to the FPGA-based data acquisition system \cite{SAHOO2021165457}. The effective run time of the detector was 48.6 days. The experimental setup at JUSL including the block diagram is shown in Fig.[\ref{fig1}]. The calibration was done with $^{241}AmBe$ (10 mCi) neutron source at a similar temperature as that of JUSL.
	
	\section{Results and Discussion}
	\label{resultsanddiscussion}
	%\lipsum[3]
	
\begin{figure}[htbp]
		\centering 
		\includegraphics[width=0.9\textwidth]{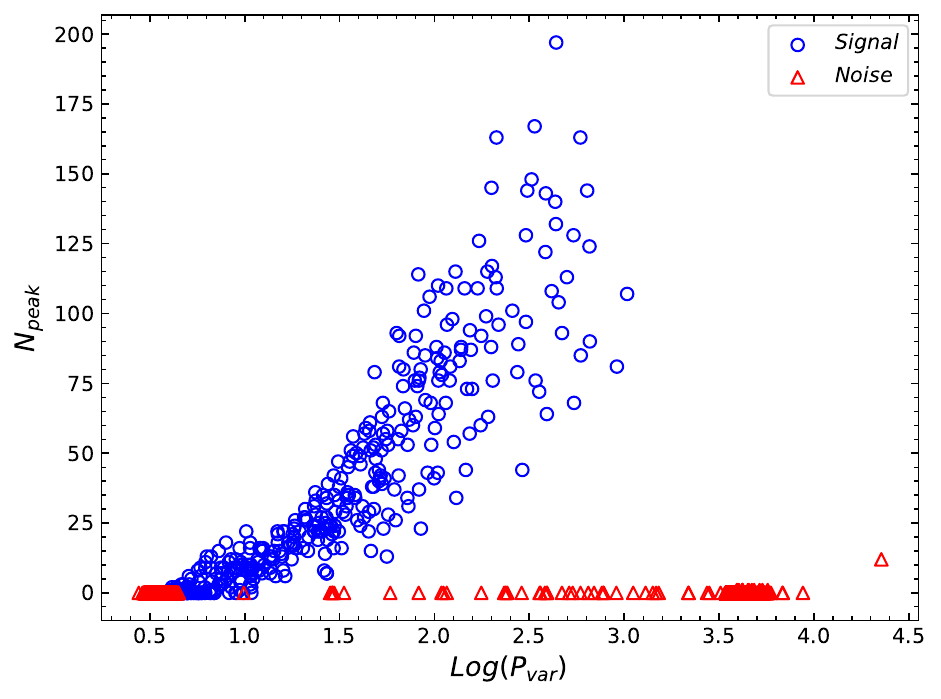}
		\caption{The distribution of the noise and the bubble nucleated signal} 
		\label{fig2}
	\end{figure}

	The collected data are in the LabVIEW file format and plotting has been done in Python code for each signal to observe the nature of the events. The signals are analyzed using the Python code by using the parameters $P_{var}$ and $N_{peak}$ where $P_{var}$ is the summation over the amplitudes of the signals for the duration of the signal and $N_{peak}$ is the number of peaks in a signal. The noises and bubble nucleation signals at JUSL are shown in Fig.[\ref{fig2}].
	
\begin{figure}[ht]
		\centering 
		\includegraphics[width=0.9\textwidth]{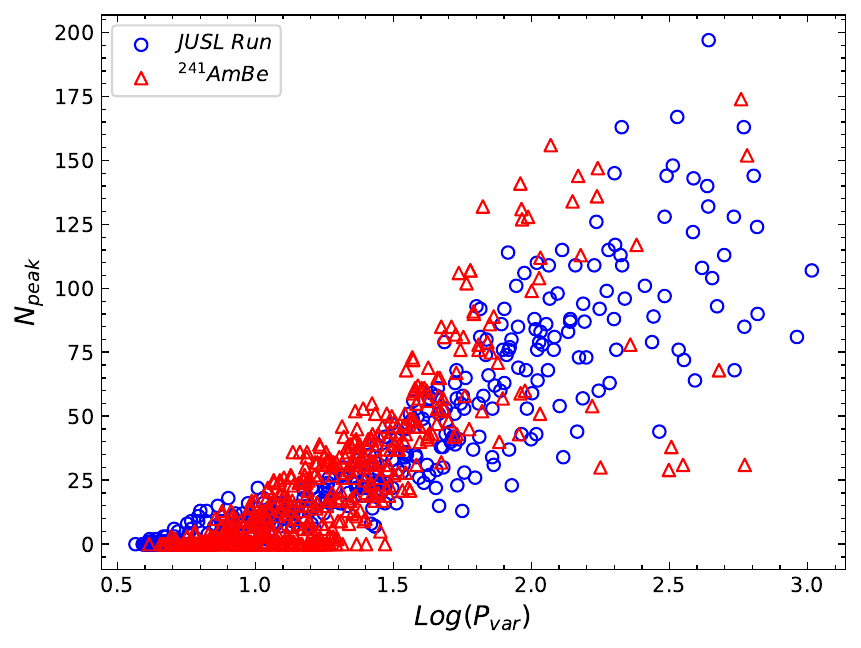}
		\caption{The $P_{var}$ and $N_{peak}$ for the calibration run and JUSL Run} 
		\label{fig3}
	\end{figure}

	It is observed that the noises are well separated from the signal in Fig.[\ref{fig2}] and in the analysis event by event separation of the noise has also been done by eye selection \cite{ALI2022166186}. The result for the calibration run and for the JUSL run is shown in Fig.[\ref{fig3}].
	
\begin{figure}[ht]
		\centering 
		\includegraphics[width=0.9\textwidth]{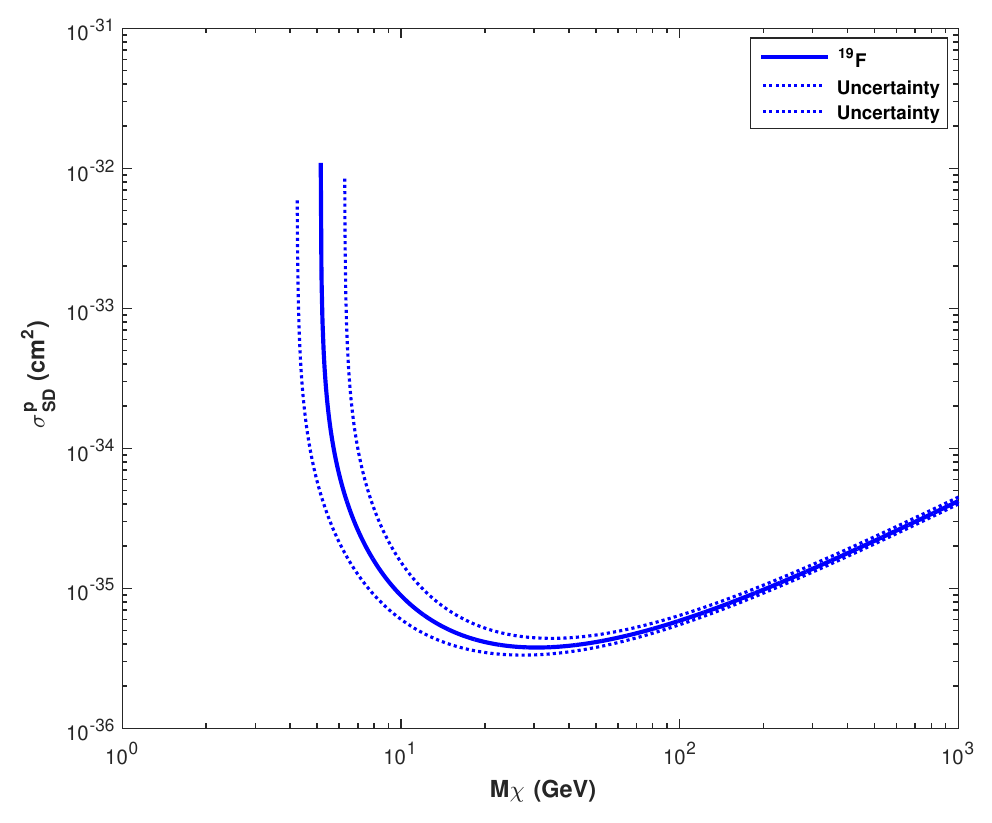}
		\caption{The upper limit of SD-p cross-section for the fluorine at 5.87 keV threshold and 2.47 kg-days of exposure from JUSL Run}
		\label{fig4}
	\end{figure}

\begin{figure}[ht]
		\centering 
		\includegraphics[width=0.9\textwidth]{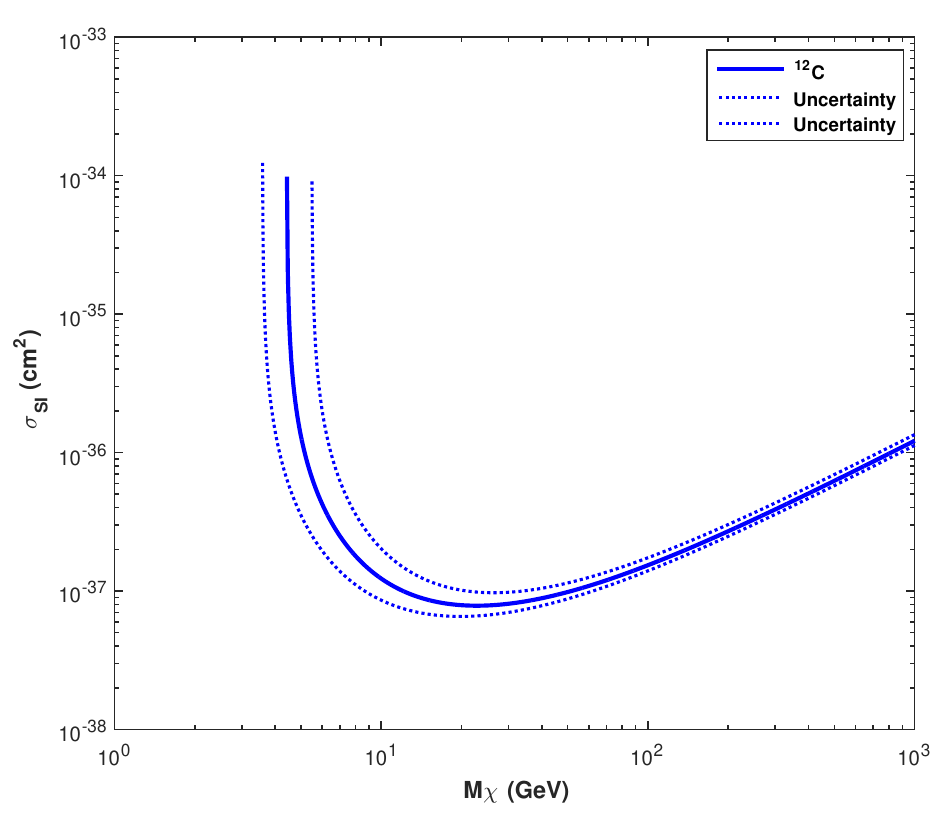}
		\caption{Upper limit of SI cross-section for carbon at 5.87 keV threshold and 2.47 kg-days of exposure from JUSL Run} 
		\label{fig5}
	\end{figure}

	The Fig.[\ref{fig3}] shows that the signals from the JUSL run are in the range of neutron-induced signals of the calibration run. The main contributing background at the 5.87 keV threshold at JUSL is the neutrons. The count rate of the detector has been estimated for the radiogenic neutrons \cite{GHOSH2022102700} as $(8.91\pm0.21)\times 10^{-6}$ /gm/sec for 100 keV-15 MeV. The contributed event from cosmogenic neutrons \cite{GHOSH2022102700} is very low which is estimated to be of the order of $10^{-11}$ /gm/sec. The present experimentally observed count rate at JUSL is $2.12\times 10^{-6}\pm9.97\times 10^{-8}$ /gm/sec which is below the count rate from the expected neutron background. It has already been observed that the $C_2H_2F_4$ SED becomes sensitive to gamma-rays at and above $38.5\pm1.4^{\circ}C$ \cite{SAHOO201944}. Therefore the events at JUSL are mainly from the background neutrons. This background limits the WIMP detection at this threshold and puts an upper limit of WIMP detection. The estimated value of the upper limit of the WIMP-nucleon cross-section for the spin-dependent (SD) interaction with fluorine nuclei and for the spin-independent (SI) interaction for the carbon nuclei are shown in Fig.[\ref{fig4}] and Fig.[\ref{fig5}] respectively. 
	
	In evaluating the exclusion plots the count rate is considered as the experimentally observed count rate which gives the upper bounds in the WIMPs detection. The efficiency for C and F is taken from the earlier published results \cite{PhysRevD.100.022001,PhysRevD.101.103005}. For the estimation of cross-section the methodology as explained in Ref \cite{PhysRevD.101.103005,LEWIN199687,Soumini_Chaudhury_2010} is considered. The minimum value of WIMP mass for both the C and F nuclei have been estimated for a 5.87 keV threshold and are shown in Table-[\ref{table1}]. 
	
	\begin{table}[!tbp]
		\caption{The threshold WIMP mass and the upper limit of cross-section from the JUSL Run, Stat: Statistical Error; Sys: Systematic Error}
		\label{table1}
		\centering
		\begin{tabular}{c c c c }
			\hline
			\hline
			Nucleus & $M_{\chi}$(min) & $M_{\chi}$(Upper Limit) & cross-section \\ 
			& $(GeV)$         & $(GeV)$                 & $(cm^2)$ \\
			\hline
			\hline
			$^{19}F$-SD & 5.16 & 30.67 & $[3.782\pm(0.179)_{stat}(^{+0.655}_{-0.432})_{sys}]\times 10^{-36}$ \\
			$^{19}F$-SI & 5.16 & 30.67 & [$7.939\pm(0.375)_{stat}(^{+1.386}_{-0.909})_{sys}]\times 10^{-39}$ \\
			$^{12}C$-SI & 4.44 & 22.81 & [$7.834\pm(0.370)_{stat}(^{+2.005}_{-1.241})_{sys}]\times 10^{-38}$ \\
			\hline
			\hline
		\end{tabular}
	\end{table}
	
	Table-[\ref{table1}] shows that upper limit of cross-section appears at 30.67 $GeV/c^2$ for $\sigma_{SD}^p$ of $[3.782\pm(0.179)_{stat}(^{+0.655}_{-0.432})_{sys}]\times 10^{-36}$ for fluorine. For the SI-interaction with carbon, the upper limit appears as [$7.834\pm(0.370)_{stat}(^{+2.005}_{-1.241})_{sys}]\times 10^{-38}$ at 22.81 $GeV/c^2$. 
	
\begin{figure}[ht]
		\centering 
		\includegraphics[width=0.9\textwidth]{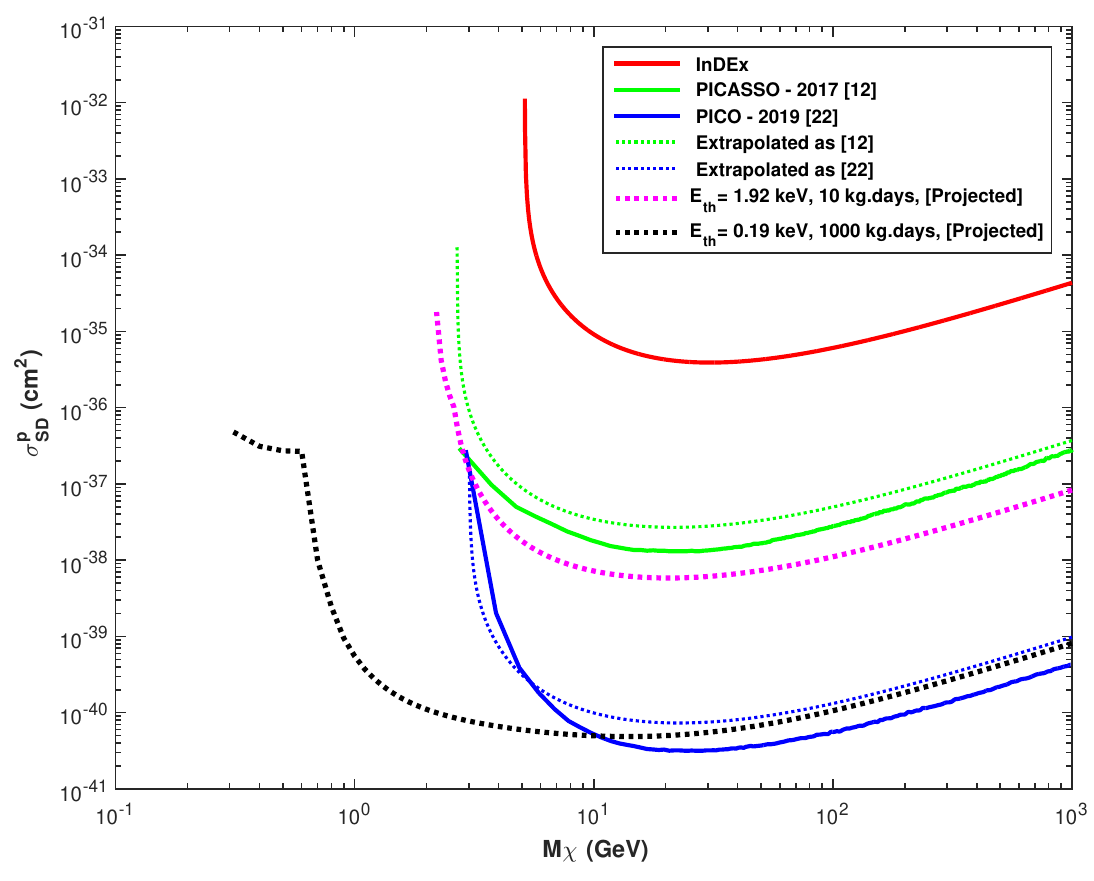}	
		\caption{Present and projected SD sensitivity of InDEx} 
		\label{fig6}
	\end{figure}

\begin{figure}[ht]
		\centering 
		\includegraphics[width=0.9\textwidth]{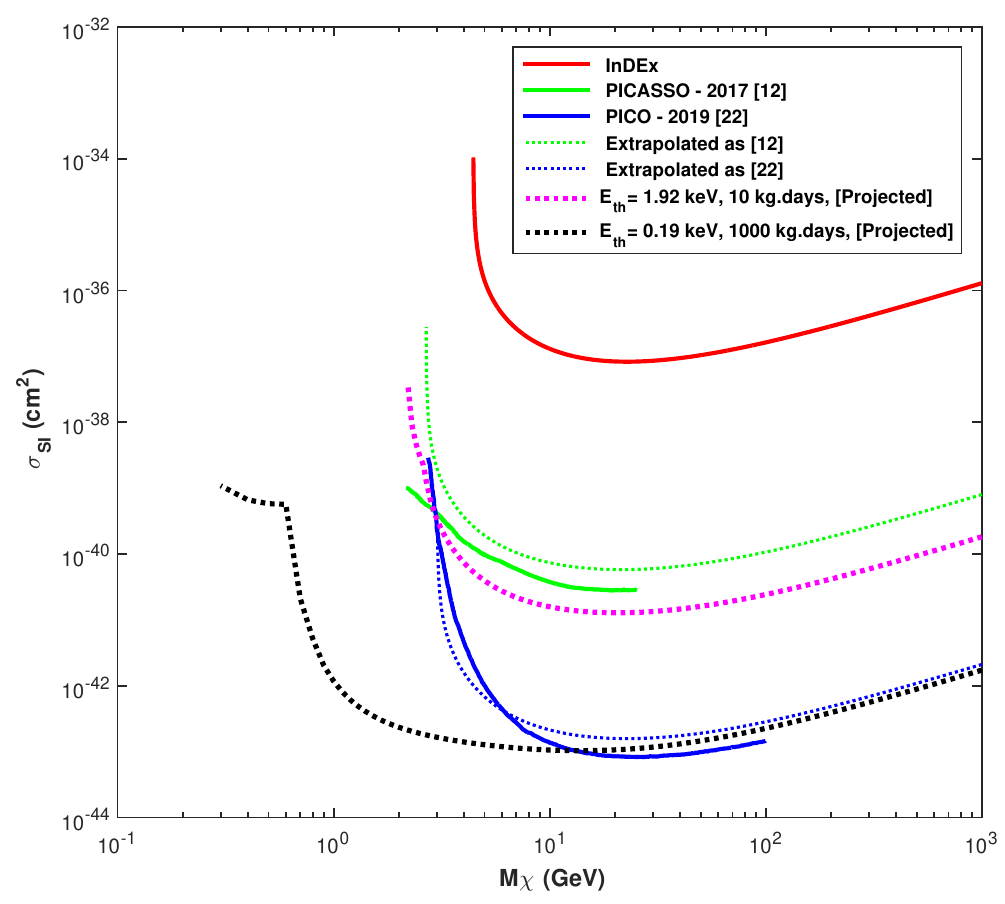}	
		\caption{Present and projected SI sensitivity of InDEx} 
		\label{fig7}
	\end{figure}

	The exclusion plots at 90\% C.L. have been generated for the present experiment and compared it with the other available results of PICASSO \cite{BEHNKE201785} and PICO \cite{PhysRevD.100.022001} experiments that use the superheated liquids, $C_4F_{10}$ and $C_3F_8$ as the targets respectively. The results are shown in Fig.[\ref{fig6}] for SD and in Fig.[\ref{fig7}] for the SI interaction. The present result has been extrapolated to the exposure and threshold of refs \cite{BEHNKE201785,PhysRevD.100.022001} and is included in the Figures. 
	
	The Fig.[\ref{fig6}] and Fig.[\ref{fig7}] show that if the present experiment is extended to the exposure and threshold of refs \cite{BEHNKE201785,PhysRevD.100.022001}, it will reach the sensitivity as shown by these two experiments. It is already shown by calculation that at a threshold of 0.19 keV, this detector can be sensitive to H-nuclei which will make it sensitive to sub-GeV WIMPs \cite{PhysRevD.101.103005}. The future projected sensitivity of InDEx for 10 kg-days at 1.92 keV threshold and for 1000 kg-days at 0.19 keV threshold have also been shown for the zero background consideration for both the SD and SI sectors.

	\section{Summary and Conclusion}
	\label{summaryandconclusion}
	The first result on the InDEx dark matter direct search for the WIMP-nucleon scattering from {\activeliquid} superheated emulsion detector at JUSL has been reported in this article. The detector ran at 5.87 keV threshold and for 2.46 kg-days of exposure. The upper limit on the spin-independent and spin-dependent cross section for the fluorine and carbon nuclei respectively have been presented at a WIMP mass of 22.81 $GeV/c^2$ and 30.67 $GeV/c^2$. The detector is sensitive to the lowest WIMP mass of 4.44 $GeV/c^2$ for C and 5.16 $GeV/c^2$ for F at the threshold of 5.87 keV. The projected sensitivity of InDEx shows promising results while operating at a lower threshold and larger exposure which will be executed in coming years.
	
	\section*{Acknowledgements}
	The authors would like to acknowledge the help and support from UCIL, Jaduguda Mine for executing the experiment at JUSL. The authors are grateful to Dr A K Mohanty, Former Director, SINP for encouraging us in setting up the DM search experiment at JUSL. The authors are also thankful to VECC for providing the $^{241}AmBe$ source. The authors are grateful to PICO Collaboration for valuable discussion. The author (V. Kumar) is thankful to CSIR
	for providing financial support for the fellowship via grant no. 09/489(0124)/2019-EMR-I.
	
	\appendix
	
	%\section{Appendix title 1}
	
	%\section{Appendix title 2}
	
	%% bibitems, please use.
	\bibliographystyle{elsarticle-num}
	\bibliography{references}
	
	%% else use the following coding to input the bibitems directly in the TeX file.
	
	%% For example:
	%% \bibitem[Aladro et al.(2015)]{Aladro15} Aladro, R., Martín, S., Riquelme, D., et al. 2015, \aas, 579, A101
	%%\end{thebibliography}
	
\end{document}